\documentclass[12pt]{iopart}

\usepackage{times}

\usepackage[english]{babel}
\usepackage[]{graphicx}

\chardef\bslash=`\\

\begin{document}


\title[]{Structural transitions of finite spherical Yukawa crystals}
\author[H. Baumgartner et al.]{H. Baumgartner, D. Asmus, V. Golubnychiy, P. Ludwig, H. K\"{a}hlert, and M. Bonitz}
\address{Institut f\"{u}r Theoretische Physik und Astrophysik, Christian-Albrechts-Universit\"{a}t, Leibnizstr. 15, 24118 Kiel}
\ead{baum@theo-physik.uni-kiel.de}

\begin{abstract}
  Small three-dimensional strongly coupled clusters of charged particles in a spherical confinement potential arrange themselves in nested concentric shells. If the particles are immersed into a background  plasma the interaction is screened. The cluster shell configuration is known to be sensitive to the screening strength. With increased screening an increased population of the inner shell(s) is observed.
Here, we present a detailed analysis of the ground state shell configurations and configuration changes in a wide range of screening parameters for clusters with particle numbers $N$ in the range of $11$ to $60$. We report three types of anomalous behaviors which are observed upon increase of screening, at fixed $N$ or for an increase of $N$ at fixed screening.
The results are obtained by means of extensive first principle molecular dynamics simulations.
\end{abstract}

\maketitle

\section{\label{sec1}Introduction}

Coulomb crystal formation is among the most exciting cooperative phenomena in charged particle systems and has been observed in a variety of fields, including ultracold ions in Paul and Penning traps \cite{wineland87,drewsen98,dubin99}, electrons and excitons in semiconductor quantum dots \cite{filinov01} and bilayers \cite{ludwig03,ludwig07}. Coulomb crystallization occurs also in classical and quantum two-component systems such as electron-ion or electron-hole plasmas \cite{bonitz05,bonitz06_2} or laser cooled expanding plasmas \cite{pohl04}, for a recent overview see \cite{bonitz08}. Of particular recent interest has been crystallization of charged microspheres in complex plasmas in two dimensions \cite{chu94,thomas94}, as well as in three dimensions \cite{arp04} since here the structure and dynamics of the individual particles is directly visible or recordable by standard CCD cameras, e.g. \cite{kaeding06}.  
\\
From the theoretical side, the shell structure of spherically confined Coulomb crystals has been analyzed in great detail by computer simulations, e.g. \cite{hasse91,tsuruta93} and references therein. More accurate data including metastable states have recently been presented \cite{ludwig05,arp05_2,golubnychiy06,apolinario07,baum07} resulting in a very good understanding of these systems. However, in dusty plasmas the interaction of the two particles forming a crystal is screened by the surrounding electrons and ions which has a significant influence on the crystal structure. In \cite{bonitz06,apolinario_phd} it was shown, by comparison with experiments, that the pair interaction is well described by an isotropic Yukawa potential and shell configurations for various values of the screening strength have been presented. It was found that screening leads to a cluster compression, a change of the average density profile \cite{henning06,henning07} and to an enhanced population of the inner shells. Yet a detailed understanding of how these shell occupation changes proceed is still missing. This is the goal of the present paper.
\\
Here we present a detailed analysis of the ground state configurations of mesoscopic clusters interacting via an isotropic Yukawa potential containing $11$ to $60$ particles in a wide range of screening parameters $0.0 \le \kappa \le 20.0$. While the general trend that with increased $\kappa$ particles move inward is confirmed, we observe several anomalies which are due to symmetry effects:
1. upon $\kappa$ increase two particles move to the inner shell at once. 2. when the particle number is increased by one at a fixed $\kappa$ one particle move from the inner to the outer shell and 3. at very large $\kappa$ there exist cases of reentrent shell fillings: one particle returns from the inner to the outer shell.

\section{\label{sec2}Model and Simulation technique}
We consider $N$ identical Yukawa interacting classical particles  with mass $m$ and charge $q$ in a three-dimensionsal isotropic harmonic confinement potential described by the hamiltonian
\begin{equation}
\label{eqn1}
	H (\textbf{r}_i, \textbf{v}_i) = \sum_i^N \frac{m}{2} \textbf{v}_i^2 + \sum_i^N \frac{\alpha}{2} \textbf{r}_i^2 + \sum_{ij}^N \frac{q^2}{4 \pi \epsilon  r_{ij}} \cdot e^{-\kappa r_{ij}}.
\end{equation}
This model has been found close to the experimental situation under which spherical dust crystals form \cite{arp05}. In the simulations we use dimensionless length and energy variables by introducing the units $r_0 = (q^2/16 \pi \epsilon \alpha)^{1/3}$ and $E_0 = (\alpha q^4/32 \pi^2 \epsilon^2)^{1/3}$, respectively.
\\
This model was already used to find the ground state configurations and their energies for Coulomb interaction in \cite{ludwig05,arp05_2}. Here we extended the investigation to the ground states of finite Yukawa systems. To obtain the ground states we perform extensive molecular dynamics simulation using a standard simulated annealing technique, e.g.\cite{kirkpatrick83}. This is done by slowing down the particles by some friction in every time step, starting from a random configuration. A stable state is reached when the dimensionless force on each particle is zero (less than $10^{-6}$ in the calculations). It was observed previously for Coulomb systems that there exist different states with the same shell configuration, which differ with respect to the particle arrangement within the shells (fine structure) \cite{ludwig05,golubnychiy06,apolinario07}. Here, these energy differences which are less than $10^{-8}$ in dimensionless units will not be resolved since this would blow up the whole analysis and we record only the energetically lowest shell configurations for a given value of $\kappa$. 

Metastable states with a different shell configuration are sometimes energetically very close to the ground state. Also, their number is increasing approximately exponentially with $N$ \cite{apolinario07} which requires special care in the numerical approach, in particular in the choice of the cooling speed. Also, for given parameters, the cooling process has to be repeated sufficiently often. In the present calculations, we typically used $10^3$-$10^4$ independent runs for every set of $(N,\kappa)$. While this does not guarantee that the true ground state is found it does ensure a sufficiently high probability that no other state with lower energy exists. As an independent tool to verify the results we performed for a number of cases standard Metropolis Monte Carlo simulations.

We simulated particle numbers from $11$ to $60$ and screenings from $\kappa = 0$ to $\kappa = 5.0$. The screening parameter was changed in steps of $\Delta \kappa = 0.1$. When for some $N$ a configuration change  at some critical $\kappa$ was detected, the calculation around this point was repeated with a substantial smaller $\kappa$ step to ensure an accuracy of $\pm 0.05$.
The choice of this interval of screening parameters is motivated by the situation of typical dusty plasma experiments where $\kappa$ is around $1$. Besides, it is of theoretical interest what will be the asymptotic shell configuration in the limit of a very short range interaction. To this end we also analyzed the ground state at $\kappa=20$ and recorded structural transitions occuring between $\kappa=5$ and $\kappa=20$.

\section{\label{sec3}Numerical results}
\subsection{\label{ssec_energy}Total energy}
\begin{figure}
	\centering
	\includegraphics[width=7.2cm]{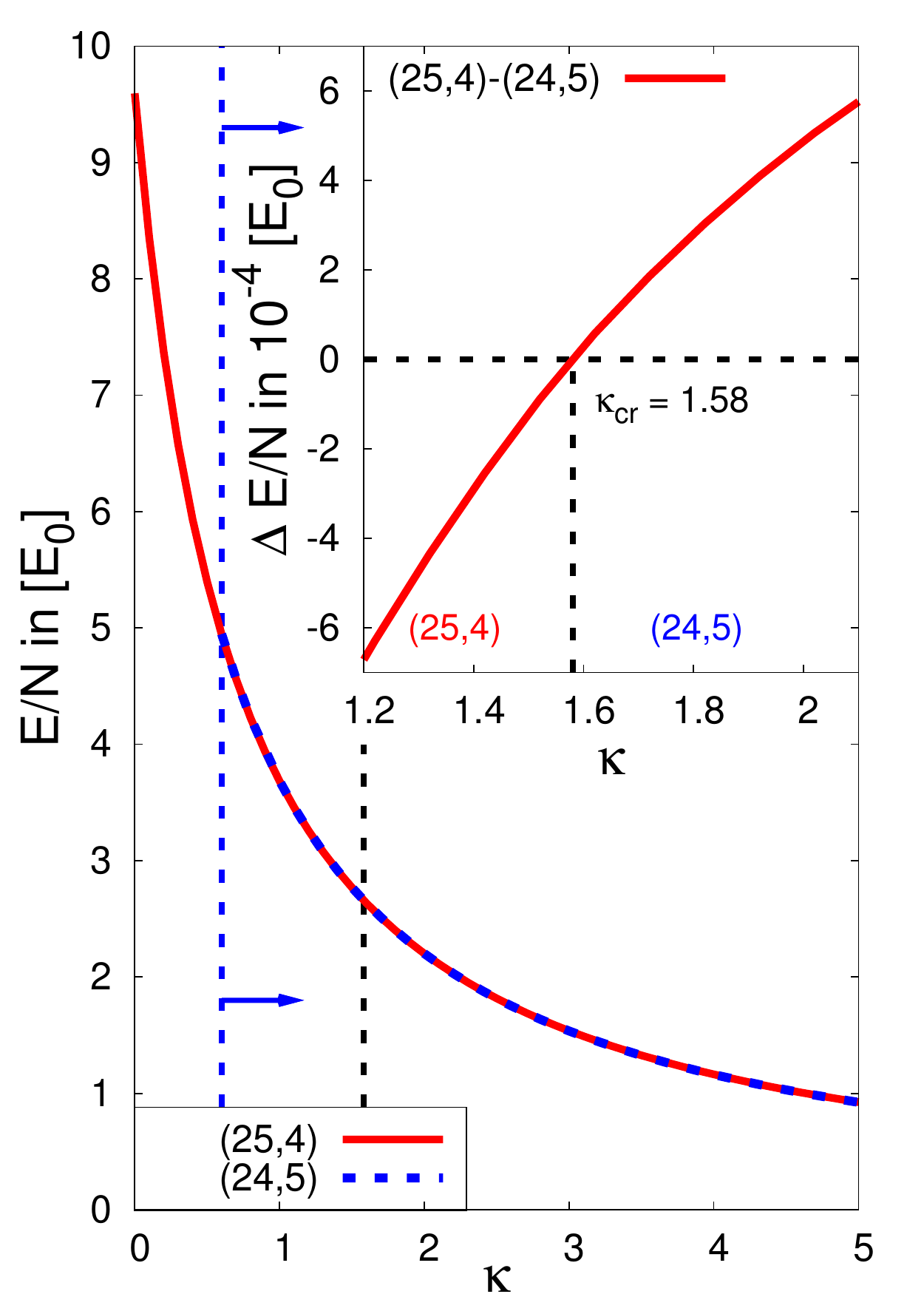}
	\caption{\label{fig2} (colour online) Energy per particle of a Yukawa cluster with $N=29$ particles for screenings $0.0 \le \kappa \le 5.0$. The red solid [blue dashed] line indicates the configuration $(25,4)$, [the configuration $(24,5)$]. The vertical blue dashed line denotes the screening from which the configuration $(24,5)$ begins to occur in the simulations. The configuration $(25,4)$ is present in the complete range of screening. The inset shows the energy difference per particle of these two configurations in a small range of screening parameters around the critical value, where the ground state shell configuration changes from $(25,4)$ to $(24,5)$. The critical value is indicated by the vertical black dashed line in the figure as well as in the inset.}
\end{figure}

A typical simulation result is shown in Fig.~\ref{fig2} where we plot the total energy per particle for the cluster $N=29$ in the range of $0.0 \le \kappa \le 5.0$. As one can see the energy decreases rapidly with $\kappa$ by approximately one order of magnitude, due to the reduction of the pair interaction strength. The same behavior is observed for other particle numbers, as shown for $N=31$ in Fig.~\ref{fig3} and, for $N=57$, in Fig.~\ref{fig4}. Due to the exponential dependence on the distance one may wonder if the energy decrease with $\kappa$ follows an exponential law as well as is the case in macroscopic one-component Yukawa plasmas, e.g. \cite{hamaguchi_3d,totsuji_3d}. 

The simplest fit for the ground state total energy per particle has the form
\begin{equation}
\label{eq2}
	\frac{E^f_{GS}(\kappa,N)}{N} = E_1(N) \cdot e^{-r_1(N) \kappa} + E_0(N)
\end{equation}
and uses three $\kappa$ independent free parameters which are functions of the particle number. In the analyzed range of $N$ this dependence is found to be close to $N^{2/3}$, for the two energies $E_0$ 
and $E_1$, whereas the effective length $r_1$ in the exponent scales approximately as $N^{1/3}$.
Using the exact results for the ground state energies per particle from the molecular dynamics simulations we obtain the following best fit for the three coefficients:
\begin{eqnarray}
\label{eq3}
	E_0(N) &=& 0.015 + 0.12 N^{2/3}, \\
	E_1(N) &=& -0.81 + 0.92 N^{2/3}, \\ 
	r_1(N) &=& 0.51 + 0.19 N^{1/3}.
\end{eqnarray}
In the Coulomb limit this fit reduces to 
\begin{equation}
\label{eq2c}
	\frac{E^f_{GS}(\kappa=0,N)}{N} = E_0(N)+E_1(N) = -0.795 + 1.04 N^{2/3}.
\end{equation}
This fit is useful to understand the main trends in the analyzed parameter range and reproduces the simulation data within several percent. Some representative examples are given in Tab.~\ref{table3}.
\begin{table}
\centering
\begin{tabular}{|c|c|r|r|r|}
	\hline
	$N$&$\kappa$&$E_{GS}/N$ (MD)&$E^f_{GS}/N$ [Eq. (\ref{eq3})]&$\Delta (\%)$\\
	\hline
	\hline
	$12$&$0.0$&$4.839$&$4.656$&$-3.8$\\
	\hline
	$12$&$4.0$&$0.685$&$0.736$&$+7.4$\\
	\hline
	$58$&$0.0$&$15.875$&$14.788$&$-6.8$\\
	\hline
	$58$&$4.0$&$1.692$&$1.902$&$+12.4$\\
	\hline
\end{tabular}
\caption{\label{table3} Ground state Energies per particle from Eq.~(\ref{eq3}), compared to the exact results from MD simulations, and the relative error $\Delta$, for some examples.}
\end{table}
Further improvements can be easily achieved using e.g. the numerical results of ref.~\cite{totsuji_3d} or the analytical expressions of ref.~\cite{cioslowski}, but this is outside the goal of the present analysis.


\subsection{\label{ssec3} Structural transitions with screening}

The presented fit for ground state total energies $E^f_{GS}$ is a continuous functions of $\kappa$ and do not immediately reveal possible changes of the shell configuration.
In fact, in many cases there co-exist several stationary states (shell configurations), the energies of which may become equal at a certain value of $\kappa$. At this point a structural transition of the ground state is observed. This can be seen in Fig.~\ref{fig2} for the cluster with $N=29$ particles. For small $\kappa$ the configuration $(25,4)$ is the ground state until at the critical value of  $\kappa_{cr} = 1.58$ the configuration $(24,5)$ has the same energy and a smaller energy beyond this point, see inset of Fig.~\ref{fig2}. Thus, if $\kappa$ crosses $\kappa_{cr}$ from below, one particle of the cluster moves from the outer to the inner shell. This ground state change is accompanied by a jump of the derivative of the exact ground state energy $dE_{GS}/d\kappa$ at $\kappa_{cr}$, so this structural transition resembles a first order phase transition.

\begin{figure}
	\centering
	\includegraphics[width=8cm]{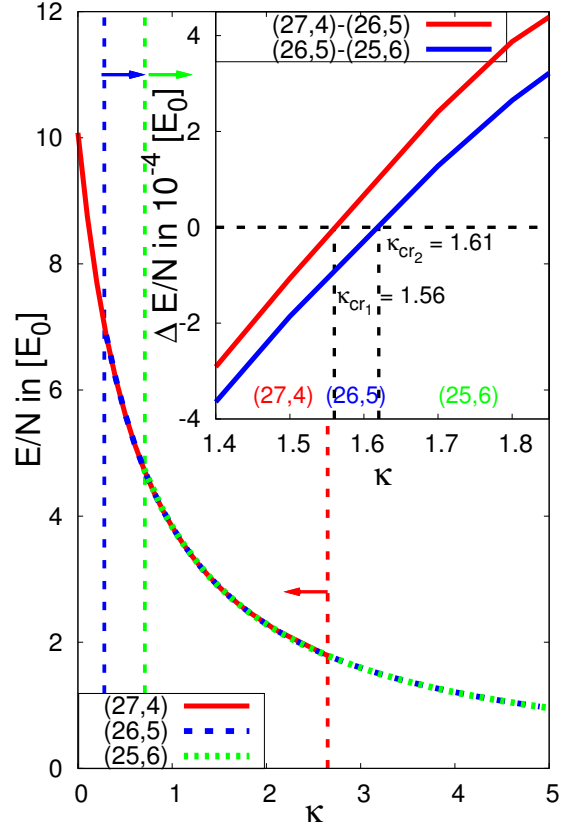}
	\caption{\label{fig3} (colour online) Energy per particle of a Yukawa cluster with $31$ particles for screenings $0.0 \le \kappa \le 5.0$. The red solid line indicates the configuration $(27,4)$ and the blue dashed [green dotted] line the configuration $(26,5)$ [$(25,6)$]. The vertical dashed lines denote the beginning [blue for $(26,5)$ and green for $(25,6)$] and the end [red for $(27,4)$] of occurance of these configurations in the simulations. The inset shows the energy difference per particle for two stable states the red [blue] solid line for the configurations $(27,4)-(26,5)$ [$(26,5)-(25,6)$] around the critical value of screening. The critical values for the changes in the ground state configurations are indicated by the vertical black dashed lines, both in the inset as well as in the figure.}
\end{figure}

Figure \ref{fig3} shows a more complicated example with two ground state changes occuring in a small range of screening parameters. For $ \kappa < 1.5623$ the ground state configuration is $(27,4)$ whereas at $\kappa_{cr1} = 1.5623$ the configuration $(26,5)$ becomes the ground state. Finally, 
at $\kappa_{cr2} = 1.6142$ this configuration is replaced by $(25,6)$ which remains the ground state for larger $\kappa$. This behavior can be seen in the energy differences plotted in the  inset of Fig.~\ref{fig3}. 
Around the interval $[\kappa_{cr1},\kappa_{cr2}]$  all three states co-exist and have very close energies which illustrates the high accuracy and fine $\kappa-$grid required in this analysis.

These two examples are typical for most cases: at small $\kappa$ the cluster structure is strongly influenced by the spherical trap. In contrast, in the limit of very large screening 
the pair interaction tends to a hard sphere interaction and the clusters approach a closed packed structure. This is often a layered structure allowing for an optimal compression \cite{baum07}. In between the two limits of long range and short range interaction the shell configurations change via one (or several) structural transitions where one particle from the outer shell moves to the inner shell as this configuration becomes energetically favorable.

There are, however, several interesting exceptions to this general behavior. We observe three kinds of ``anomalies'' which will be analyzed in the following

\subsection{\label{ssec_ano1} Anomalies of first kind: Correlated two-particle transitions}
\begin{figure}
	\centering
	\includegraphics[width=10cm]{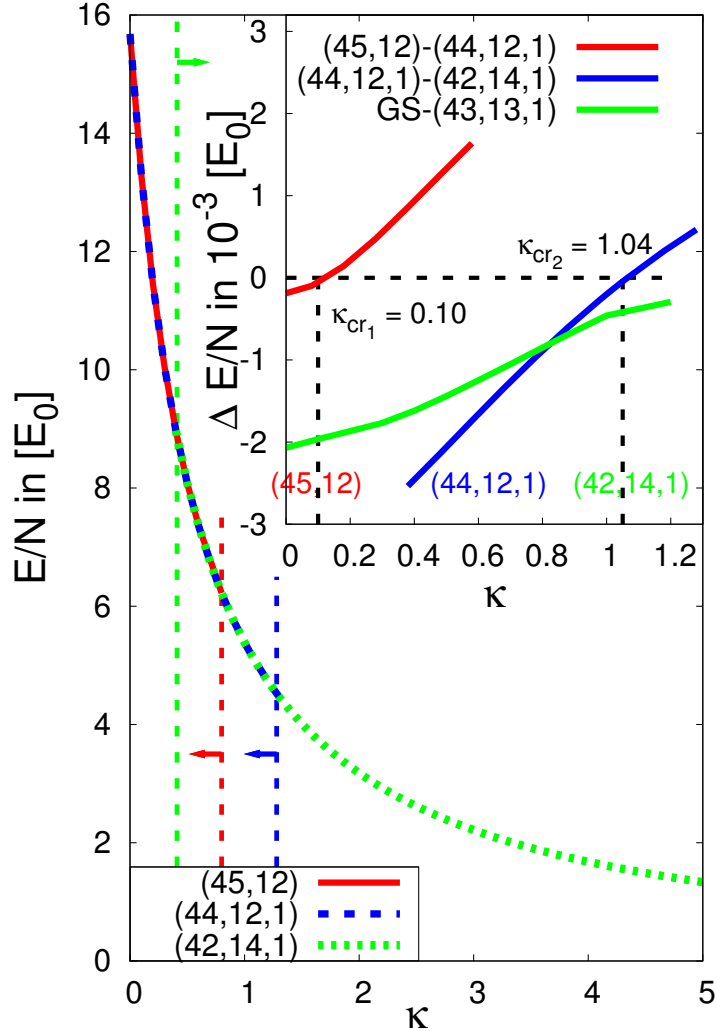}
	\caption{\label{fig4} (colour online) Energy per particle of a Yukawa cluster with $57$ particles for screenings $0.0 \le \kappa \le 5.0$. The red solid line indicates the configuration $(45,12)$ and the blue dashed [green dotted] line the configuration $(44,12,1)$ [$(42,14,1)$]. The vertical dashed lines denote the beginning [green for $(42,14,1)$)] and the end [red for $(45,12)$ and blue for $(44,12,1)$, respectively] of occurance of these configurations in the simulations. The inset shows the energy difference per particle: the red [blue] solid line for the configurations $(45,12)-(44,12,1)$ [$(44,12,1)-(42,14,1)$] around the critical range of screening. The green solid line is the energy difference of the metastable configuration $(43,13,1)$ to the current ground state, this configuration is never the ground state. The critical values for the changes in the ground state configurations are indicated by the vertical black dashed lines. The change $(44,12,1) \rightarrow (42,14,1)$ at $\kappa_{cr2} = 1.04$ shows an {\em anomaly of the first kind}.}
\end{figure}

Consider now the cluster $N=57$, cf. Fig.~\ref{fig4}. At small screening, the configuration $(45,12)$ is the ground state until at $\kappa_{cr1} = 0.10$ one particle from the outer shell moves to the cluster center forming a new shell with the configuration $(44,12,1)$. Thereby the second shell is not changed since it has a ``closed shell'' configuration with $12$ particles. Besides this ``normal'' transition, at $\kappa_{cr2} = 1.04$ a new type of structural change is observed: the configuration changes according to $(44,12,1) \longrightarrow (42,14,1)$. This means, at this point a {\em correlated intershell transition of two particles} is observed. This unusual behavior will be called ``anomaly of first kind''. The reason of this anomaly is the particularly high stability of the closed shell configuration of shell two which dominates the structure up to rather large screening. In contrast, a configuration with $13$ particles on the second shell is energetically very unfavorable, although it exists in a broad range of $\kappa$ values, in fact, the configuration $(45,13,1)$ is never the ground state as can be seen in the inset of Fig.~\ref{fig4}.

The first occurence of an anomaly of the first kind is at $N=30$ where a transition $(26,4) \rightarrow (24,6)$ is observed at $\kappa\approx 1.5$. There is a total of $18$ occurences of such anomalies: at $N=30, 34, 36, 38, 40, 45-54, 57, 58, 60$. The reason for this behaviour is that in all cases but for $N=57,58,60$ the new ground state configuration, e.g. $(24,6)$ at screening above $\kappa = 1.5$, always forms a platonic body on the inner shell. This is a highly symmetric configuration which obviously decreases the energy per particle better than by just adding one particle \cite{ludwig05,kamimura07}. For the cases $N=30,34,36,38,40$ the ground state configuration even change from one platonic body to another, while for the cases $N=46-54$ the system changes from $10$ particles on the inner shell to the closed shell configuration with $12$ particles. 

\begin{figure}
	\centering
	\includegraphics[width=15cm]{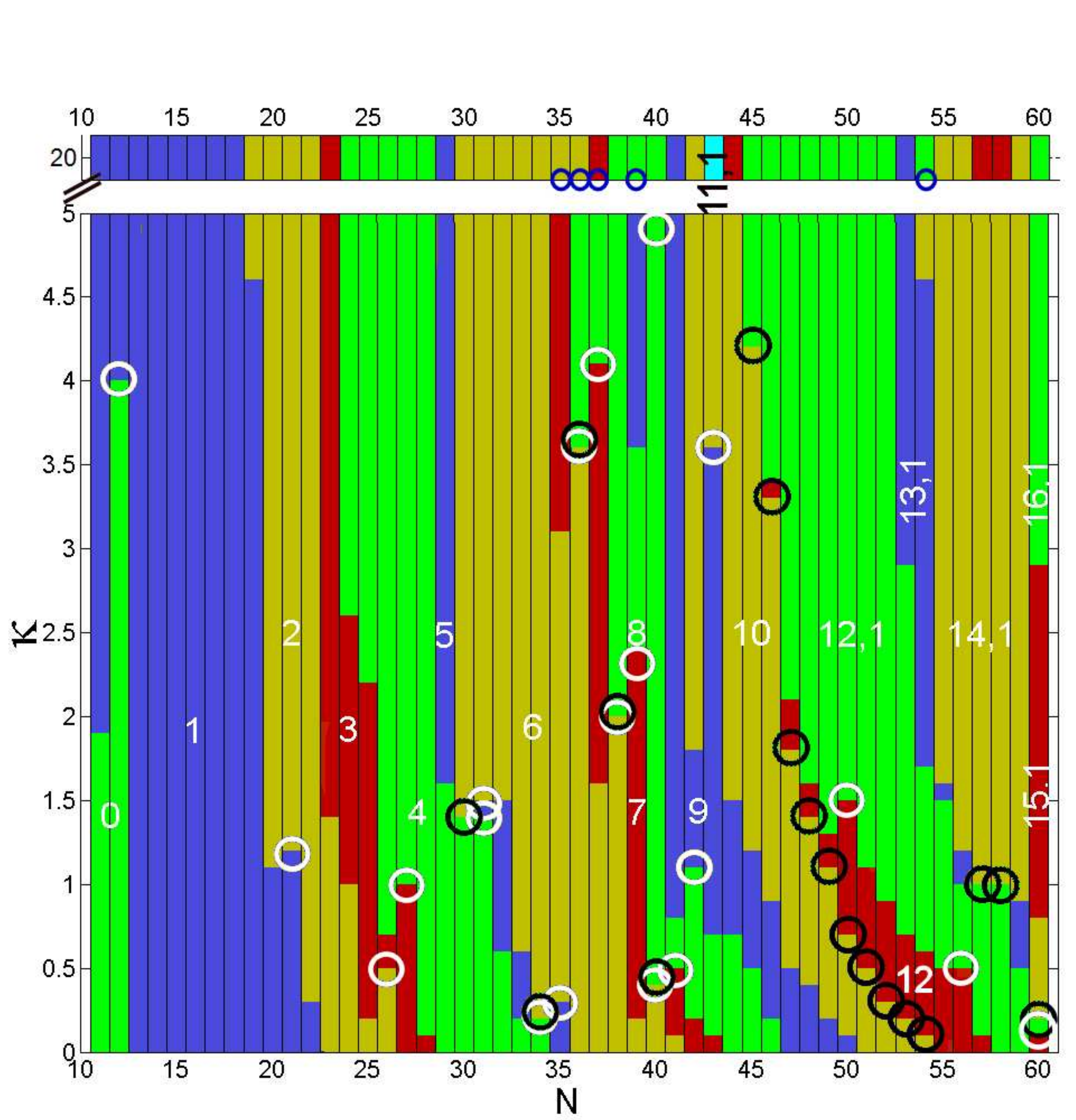}
	\caption{\label{fig1}(colour online) Ground states of small $(11\le N \le 60)$ Yukawa balls for the range of screening parameter $(0.0 \le \kappa \le 5.0)$. The numbers on the bars denote the number of particles on the inner shell(s). The black circles indicate anomalies of the 1st kind. The white circles indicate the end of the screening range, where anomalies of the 2nd kind appear. The ground states for a screening parameter $\kappa = 20.0$ are plotted above the diagram for comparison in what range the ground states at $\kappa = 5.0$ are stable. The cyan bar for $N=43$ at $\kappa = 20.0$ refers to a ground state of $(11,1)$ in the center region; it is the only time this configuration is part of a ground state. The dark blue circles just below $\kappa = 20.0$ indicate anomalies of the 3rd kind, where a ground state configuration reappears with increased screening.}
\end{figure}

The only configuration with $11$ particles on an inner shell is found for the case $N=43$, at the very large screening value of $\kappa = 20.0$, which leads to the conclusion that this configuration is energetically unfavorable. In the other three cases, $N=57,58,60$, the ground state configuration changes from $(12,1)$ to $(14,1)$ on the inner shells. Although a ground state configuration with $(13,1)$ particles in the cluster center is observed for some particle numbers in a certain range of screening parameters, the configurations $(12,1)$ and $(14,1)$ are far more often the ground state. 

These anomalies are shown in the full ground state diagram, Fig.~\ref{fig1}, by the black circles.
The complete list is also shown in Table~\ref{table1} by the bold numbers.

\subsection{\label{ssec_ano2} Anomalies of the second kind: Reduction of inner shell population upon increase of $N$}
\begin{figure}
	\centering
	\includegraphics[width=15.0cm]{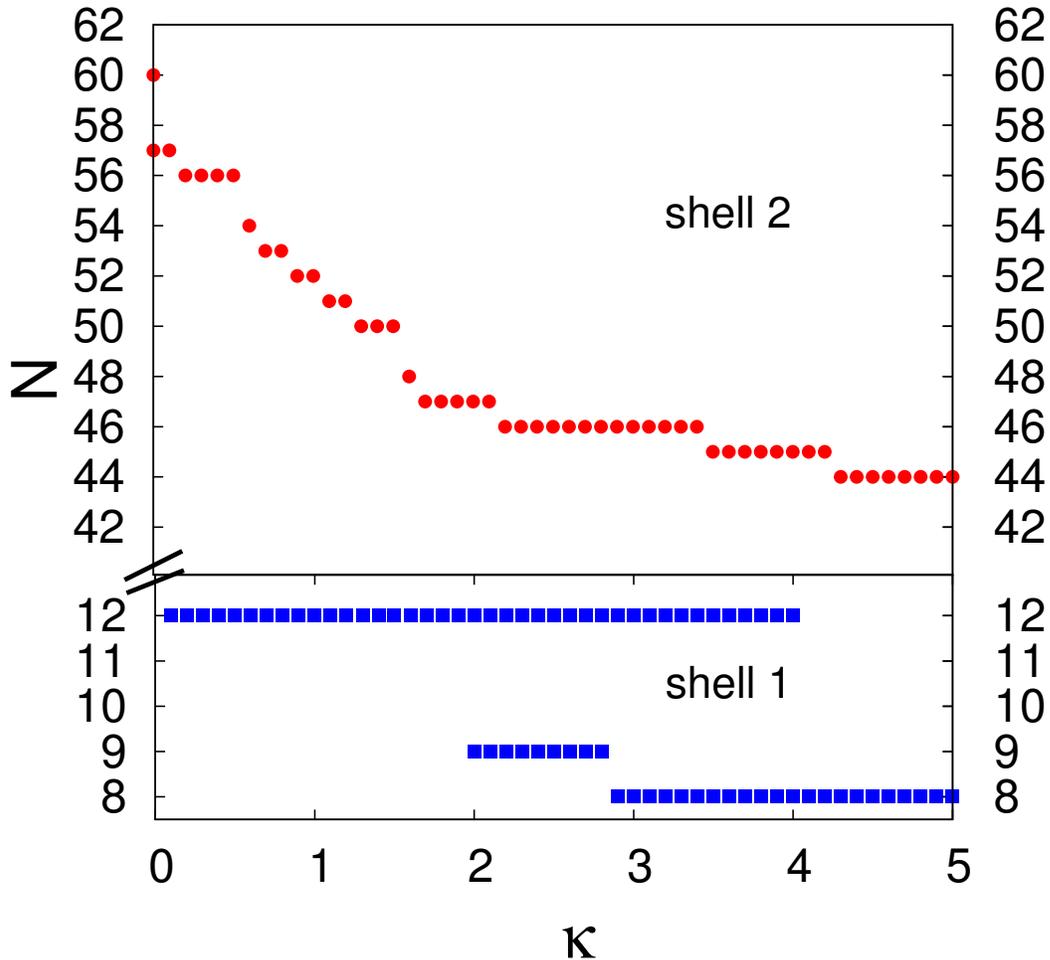}
	\caption{\label{fig8} (colour online) Shell closures for the ground state configurations for the first two shells in the range of screening $0.0 \le \kappa \le 5.0$. The particle number $N$ for the last closed shell is given by red dots [blue squares] for the shell $1$ [$2$] in the considered range of screening. 
In some cases ($\kappa = 0$ [$2.0 \le \kappa \le 4.0$] for shell 2 [shell 1] there exist different number of particles with closed shells, e.g. $N=57$ and $N=60$ for $\kappa = 0$.}
\end{figure}

Let us now consider changes of the total particle number $N$ at constant screening. The ``normal'' trend upon an increase of the particle number by one is, of course, that the new particle is added to one of the existing shells (leaving the other shells unchanged) or moves into the center opening a new shell. However, again, one observes exceptions from this rule, cf. Fig.~\ref{fig1}. This effect was already observed for the Coulomb cluster ($\kappa=0$) with $N=59$ \cite{ludwig05}. It has the ground state configuration $(46,12,1)$. Addition of another particle to the cluster gives rise to the configuration $(48,12)$. This is again a structural transition involving correlated behavior of two particles which we call ``Anomaly of the second kind''. In this particular case this transition is even associated with a change of the number of shells:
 the three-shell configuration [first appearing at $N=58$] disappears again and, instead, a two-shell configuration is restored. This is, of course, a consequence of the particular stability of the latter which contains two closed shells with $12$ and $48$ particles, respectively. The closed shell configurations in dependence of the screening are given in Fig.~\ref{fig8}.
 
While, in Coulomb systems, $N=59$ is the only known case of an anomaly of the second kind, in Yukawa clusters this behavior appears quite frequently.
The first occurence is at $N=11$ for $\kappa$ values between about $2$ and $4$. Here, addition of a particle gives rise to the configuration change $(10,1)\longrightarrow (12,0)$, i.e. one particle moves away from the inner shell (the shell vanishes), and the population of the outer shell increases by two. There is a total of $20$ such anomalous transitions observed for $18$ particle numbers:
$N=11, 20, 25, 26, 30, 33-42, 49, 55, 60$. There are two particle numbers where this effect occurs two times: for $N=30 \longrightarrow 31$, in the $\kappa$ range $[1.5623,1.6142]$ the configuration changes from $(24,6)$ to $(26,5)$. Interestingly, for screening parameters just below this range, i.e. $[1.4866, 1.5623]$ the inner shell loses even two particles, i.e. we observe the transition 
$(24,6)\longrightarrow (27,4)$. The second case where two such transitions occur is 
the transition $39 \longrightarrow 40$. There for $\kappa$ between $0.2223$ and $0.4179$ the ground state changes according to $(32,7)\longrightarrow (34,6)$ whereas at $\kappa > 3.612$ the configuration change is $(31,9)\longrightarrow (32,8)$.

Finally, anomalies of the second kind which are additionally associated with vanishing of one ``shell'' (i.e. removal of one particle from the cluster center) are found 4 times: for $N=11\longrightarrow 12$ the transition $(10,1)\longrightarrow (12)$ is observed (see above). Return to a two shell configuration occurs three times: for $N=49\longrightarrow 50$ we find the transition $(36,12,1)\longrightarrow (38,12)$, for $N=55\longrightarrow 56$, the transition $(42,12,1)\longrightarrow (44,12)$ and, for $N=59\longrightarrow 60$, the transition $(46,12,1)\longrightarrow (48,12)$ which is known from the Coulomb case (see above) and appears here in a narrow range of small $\kappa$ values.
The complete set of these anomalies is given in table \ref{table2}. 

\begin{table}
\small
\begin{tabular}{|l|l|l|l|l|l|l|l|l|l|l|l|}
\hline
$N$ & $\kappa_{cr}$ & GS & $N$ & $\kappa_{cr}$ & GS & $N$ & $\kappa_{cr}$ & GS & $N$ & $\kappa_{cr}$ & GS \\ 
\hline
\multicolumn{1}{|c|}{$11$} & \multicolumn{1}{r|}{$1.9$} & \multicolumn{1}{c|}{$(11,0)$} & \multicolumn{1}{c|}{$\textbf{30}$} & \multicolumn{1}{r|}{$\textbf{1.5}$} & \multicolumn{1}{c|}{$\textbf{(26,4)}$} & \multicolumn{1}{c|}{$42$} & \multicolumn{1}{r|}{$0.2$} & \multicolumn{1}{c|}{$(35,7)$} & \multicolumn{1}{c|}{$\textbf{51}$} & \multicolumn{1}{r|}{$\textbf{0.5}$} & \multicolumn{1}{c|}{$\textbf{(41,10)}$} \\ 
\multicolumn{1}{|c|}{} & \multicolumn{1}{r|}{$>5.0$} & \multicolumn{1}{c|}{$(10,1)$} & \multicolumn{1}{c|}{} & \multicolumn{1}{r|}{$>5.0$} & \multicolumn{1}{c|}{$(24,6)$} & \multicolumn{1}{c|}{} & \multicolumn{1}{r|}{$1.1$} & \multicolumn{1}{c|}{$(34,8)$} & \multicolumn{1}{c|}{} & \multicolumn{1}{r|}{$1.1$} & \multicolumn{1}{c|}{$(39,12)$} \\ 
\cline{1-6}
\multicolumn{1}{|c|}{$12$} & \multicolumn{1}{r|}{$4.1$} & \multicolumn{1}{c|}{$(12,0)$} & \multicolumn{1}{c|}{$31$} & \multicolumn{1}{r|}{$1.56$} & \multicolumn{1}{c|}{$(27,4)$} & \multicolumn{1}{c|}{} & \multicolumn{1}{r|}{$1.8$} & \multicolumn{1}{c|}{$(33,9)$} & \multicolumn{1}{c|}{} & \multicolumn{1}{r|}{$>5.0$} & \multicolumn{1}{c|}{$(38,12,1)$} \\ 
\cline{10-12}
\multicolumn{1}{|c|}{} & \multicolumn{1}{r|}{$>5.0$} & \multicolumn{1}{c|}{$(11,1)$} & \multicolumn{1}{c|}{} & \multicolumn{1}{r|}{$1.61$} & \multicolumn{1}{c|}{$(26,5)$} & \multicolumn{1}{c|}{} & \multicolumn{1}{r|}{$>5.0$} & \multicolumn{1}{c|}{$(32,10)$} & \multicolumn{1}{c|}{$\textbf{52}$} & \multicolumn{1}{r|}{$\textbf{0.3}$} & \multicolumn{1}{c|}{$\textbf{(42,10)}$} \\ 
\cline{1-3}\cline{7-9}
\multicolumn{1}{|c|}{$13$} & \multicolumn{1}{r|}{$>5.0$} & \multicolumn{1}{c|}{$(12,1)$} & \multicolumn{1}{c|}{} & \multicolumn{1}{r|}{$>5.0$} & \multicolumn{1}{c|}{$(25,6)$} & \multicolumn{1}{c|}{$43$} & \multicolumn{1}{r|}{$0.7$} & \multicolumn{1}{c|}{$(35,8)$} & \multicolumn{1}{c|}{} & \multicolumn{1}{r|}{$0.9$} & \multicolumn{1}{c|}{$(40,12)$} \\ 
\cline{1-6}
\multicolumn{1}{|c|}{$14$} & \multicolumn{1}{r|}{$>5.0$} & \multicolumn{1}{c|}{$(13,1)$} & \multicolumn{1}{c|}{$32$} & \multicolumn{1}{r|}{$0.6$} & \multicolumn{1}{c|}{$(28,4)$} & \multicolumn{1}{c|}{} & \multicolumn{1}{r|}{$3.6$} & \multicolumn{1}{c|}{$(34,9)$} & \multicolumn{1}{c|}{} & \multicolumn{1}{r|}{$>5.0$} & \multicolumn{1}{c|}{$(39,12,1)$} \\ 
\cline{1-3}\cline{10-12}
\multicolumn{1}{|c|}{$15$} & \multicolumn{1}{r|}{$>5.0$} & \multicolumn{1}{c|}{$(14,1)$} & \multicolumn{1}{c|}{} & \multicolumn{1}{r|}{$1.5$} & \multicolumn{1}{c|}{$(27,5)$} & \multicolumn{1}{c|}{} & \multicolumn{1}{r|}{$>5.0$} & \multicolumn{1}{c|}{$(33,10)$} & \multicolumn{1}{c|}{$\textbf{53}$} & \multicolumn{1}{r|}{$\textbf{0.2}$} & \multicolumn{1}{c|}{$\textbf{(43,10)}$} \\ 
\cline{1-3}\cline{7-9}
\multicolumn{1}{|c|}{$16$} & \multicolumn{1}{r|}{$>5.0$} & \multicolumn{1}{c|}{$(15,1)$} & \multicolumn{1}{c|}{} & \multicolumn{1}{r|}{$>5.0$} & \multicolumn{1}{c|}{$(26,6)$} & \multicolumn{1}{c|}{$44$} & \multicolumn{1}{r|}{$0.7$} & \multicolumn{1}{c|}{$(36,8)$} & \multicolumn{1}{c|}{} & \multicolumn{1}{r|}{$0.7$} & \multicolumn{1}{c|}{$(41,12)$} \\ 
\cline{1-6}
\multicolumn{1}{|c|}{$17$} & \multicolumn{1}{r|}{$>5.0$} & \multicolumn{1}{c|}{$(16,1)$} & \multicolumn{1}{c|}{$33$} & \multicolumn{1}{r|}{$0.2$} & \multicolumn{1}{c|}{$(29,4)$} & \multicolumn{1}{c|}{} & \multicolumn{1}{r|}{$1.5$} & \multicolumn{1}{c|}{$(35,9)$} & \multicolumn{1}{c|}{} & \multicolumn{1}{r|}{$2.9$} & \multicolumn{1}{c|}{$(40,12,1)$} \\ 
\cline{1-3}
\multicolumn{1}{|c|}{$18$} & \multicolumn{1}{r|}{$>5.0$} & \multicolumn{1}{c|}{$(17,1)$} & \multicolumn{1}{c|}{} & \multicolumn{1}{r|}{$0.6$} & \multicolumn{1}{c|}{$(28,5)$} & \multicolumn{1}{c|}{} & \multicolumn{1}{r|}{$>5.0$} & \multicolumn{1}{c|}{$(34,10)$} & \multicolumn{1}{c|}{} & \multicolumn{1}{r|}{$>5.0$} & \multicolumn{1}{c|}{$(39,13,1)$} \\ 
\cline{1-3}\cline{7-12}
\multicolumn{1}{|c|}{$19$} & \multicolumn{1}{r|}{$4.6$} & \multicolumn{1}{c|}{$(18,1)$} & \multicolumn{1}{c|}{} & \multicolumn{1}{r|}{$>5.0$} & \multicolumn{1}{c|}{$(27,6)$} & \multicolumn{1}{c|}{$\textbf{45}$} & \multicolumn{1}{r|}{$0.5$} & \multicolumn{1}{c|}{$(37,8)$} & \multicolumn{1}{c|}{$\textbf{54}$} & \multicolumn{1}{r|}{$\textbf{0.0}$} & \multicolumn{1}{c|}{$\textbf{(44,10)}$} \\ 
\cline{4-6}
\multicolumn{1}{|c|}{} & \multicolumn{1}{r|}{$>5.0$} & \multicolumn{1}{c|}{$(17,2)$} & \multicolumn{1}{c|}{$\textbf{34}$} & \multicolumn{1}{r|}{$\textbf{0.2}$} & \multicolumn{1}{c|}{$\textbf{(30,4)}$} & \multicolumn{1}{c|}{} & \multicolumn{1}{r|}{$1.2$} & \multicolumn{1}{c|}{$(36,9)$} & \multicolumn{1}{c|}{} & \multicolumn{1}{r|}{$0.6$} & \multicolumn{1}{c|}{$(42,12)$} \\ 
\cline{1-3}
\multicolumn{1}{|c|}{$20$} & \multicolumn{1}{r|}{$1.1$} & \multicolumn{1}{c|}{$(19,1)$} & \multicolumn{1}{c|}{} & \multicolumn{1}{r|}{$>5.0$} & \multicolumn{1}{c|}{$(28,6)$} & \multicolumn{1}{c|}{} & \multicolumn{1}{r|}{$\textbf{4.2}$} & \multicolumn{1}{c|}{$\textbf{(35,10)}$} & \multicolumn{1}{c|}{} & \multicolumn{1}{r|}{$1.7$} & \multicolumn{1}{c|}{$(41,12,1)$} \\ 
\cline{4-6}
\multicolumn{1}{|c|}{} & \multicolumn{1}{r|}{$>5.0$} & \multicolumn{1}{c|}{$(18,2)$} & \multicolumn{1}{c|}{$35$} & \multicolumn{1}{r|}{$0.3$} & \multicolumn{1}{c|}{$(30,5)$} & \multicolumn{1}{c|}{} & \multicolumn{1}{r|}{$>5.0$} & \multicolumn{1}{c|}{$(32,12,1)$} & \multicolumn{1}{c|}{} & \multicolumn{1}{r|}{$>5.0$} & \multicolumn{1}{c|}{$(40,13,1)$} \\ 
\cline{1-3}\cline{7-12}
\multicolumn{1}{|c|}{$21$} & \multicolumn{1}{r|}{$1.2$} & \multicolumn{1}{c|}{$(20,1)$} & \multicolumn{1}{c|}{} & \multicolumn{1}{r|}{$3.2$} & \multicolumn{1}{c|}{$(29,6)$} & \multicolumn{1}{c|}{$\textbf{46}$} & \multicolumn{1}{r|}{$0.2$} & \multicolumn{1}{c|}{$(38,8)$} & \multicolumn{1}{c|}{$55$} & \multicolumn{1}{r|}{$0.5$} & \multicolumn{1}{c|}{$(43,12)$} \\ 
\multicolumn{1}{|c|}{} & \multicolumn{1}{r|}{$>5.0$} & \multicolumn{1}{c|}{$(19,2)$} & \multicolumn{1}{c|}{} & \multicolumn{1}{r|}{$>5.0$} & \multicolumn{1}{c|}{$(28,7)$} & \multicolumn{1}{c|}{} & \multicolumn{1}{r|}{$0.9$} & \multicolumn{1}{c|}{$(37,9)$} & \multicolumn{1}{c|}{} & \multicolumn{1}{r|}{$1.5$} & \multicolumn{1}{c|}{$(42,12,1)$} \\ 
\cline{1-6}
\multicolumn{1}{|c|}{$22$} & \multicolumn{1}{r|}{$0.3$} & \multicolumn{1}{c|}{$(21,1)$} & \multicolumn{1}{c|}{$\textbf{36}$} & \multicolumn{1}{r|}{$\textbf{3.6}$} & \multicolumn{1}{c|}{$\textbf{(30,6)}$} & \multicolumn{1}{c|}{} & \multicolumn{1}{r|}{$\textbf{3.3}$} & \multicolumn{1}{c|}{$\textbf{(36,10)}$} & \multicolumn{1}{c|}{} & \multicolumn{1}{r|}{$1.6$} & \multicolumn{1}{c|}{$(41,13,1)$} \\ 
\multicolumn{1}{|c|}{} & \multicolumn{1}{r|}{$>5.0$} & \multicolumn{1}{c|}{$(20,2)$} & \multicolumn{1}{c|}{} & \multicolumn{1}{r|}{$>5.0$} & \multicolumn{1}{c|}{$(28,8)$} & \multicolumn{1}{c|}{} & \multicolumn{1}{r|}{$3.4$} & \multicolumn{1}{c|}{$(34,12)$} & \multicolumn{1}{c|}{} & \multicolumn{1}{r|}{$>5.0$} & \multicolumn{1}{c|}{$(40,14,1)$} \\ 
\cline{1-6}\cline{10-12}
\multicolumn{1}{|c|}{$23$} & \multicolumn{1}{r|}{$1.4$} & \multicolumn{1}{c|}{$(21,2)$} & \multicolumn{1}{c|}{$37$} & \multicolumn{1}{r|}{$1.7$} & \multicolumn{1}{c|}{$(31,6)$} & \multicolumn{1}{c|}{} & \multicolumn{1}{r|}{$>5.0$} & \multicolumn{1}{c|}{$(33,12,1)$} & \multicolumn{1}{c|}{$56$} & \multicolumn{1}{r|}{$0.5$} & \multicolumn{1}{c|}{$(44,12)$} \\ 
\cline{7-9}
\multicolumn{1}{|c|}{} & \multicolumn{1}{r|}{$>5.0$} & \multicolumn{1}{c|}{$(20,3)$} & \multicolumn{1}{c|}{} & \multicolumn{1}{r|}{$4.2$} & \multicolumn{1}{c|}{$(30,7)$} & \multicolumn{1}{c|}{$\textbf{47}$} & \multicolumn{1}{r|}{$0.5$} & \multicolumn{1}{c|}{$(38,9)$} & \multicolumn{1}{c|}{} & \multicolumn{1}{r|}{$1.0$} & \multicolumn{1}{c|}{$(43,12,1)$} \\ 
\cline{1-3}
\multicolumn{1}{|c|}{$24$} & \multicolumn{1}{r|}{$1.0$} & \multicolumn{1}{c|}{$(22,2)$} & \multicolumn{1}{c|}{} & \multicolumn{1}{r|}{$>5.0$} & \multicolumn{1}{c|}{$(29,8)$} & \multicolumn{1}{c|}{} & \multicolumn{1}{r|}{$\textbf{1.8}$} & \multicolumn{1}{c|}{$\textbf{(37,10)}$} & \multicolumn{1}{c|}{} & \multicolumn{1}{r|}{$1.2$} & \multicolumn{1}{c|}{$(42,13,1)$} \\ 
\cline{4-6}
\multicolumn{1}{|c|}{} & \multicolumn{1}{r|}{$2.6$} & \multicolumn{1}{c|}{$(21,3)$} & \multicolumn{1}{c|}{$\textbf{38}$} & \multicolumn{1}{r|}{$\textbf{2.0}$} & \multicolumn{1}{c|}{$\textbf{(32,6)}$} & \multicolumn{1}{c|}{} & \multicolumn{1}{r|}{$2.1$} & \multicolumn{1}{c|}{$(35,12)$} & \multicolumn{1}{c|}{} & \multicolumn{1}{r|}{$>5.0$} & \multicolumn{1}{c|}{$(41,14,1)$} \\ 
\cline{10-12}
\multicolumn{1}{|c|}{} & \multicolumn{1}{r|}{$>5.0$} & \multicolumn{1}{c|}{$(20,4)$} & \multicolumn{1}{c|}{} & \multicolumn{1}{r|}{$>5.0$} & \multicolumn{1}{c|}{$(30,8)$} & \multicolumn{1}{c|}{} & \multicolumn{1}{r|}{$>5.0$} & \multicolumn{1}{c|}{$(34,12,1)$} & \multicolumn{1}{c|}{$\textbf{57}$} & \multicolumn{1}{r|}{$0.1$} & \multicolumn{1}{c|}{$(45,12)$} \\ 
\cline{1-9}
\multicolumn{1}{|c|}{$25$} & \multicolumn{1}{r|}{$0.3$} & \multicolumn{1}{c|}{$(23,2)$} & \multicolumn{1}{c|}{$39$} & \multicolumn{1}{r|}{$0.2$} & \multicolumn{1}{c|}{$(33,6)$} & \multicolumn{1}{c|}{$\textbf{48}$} & \multicolumn{1}{r|}{$0.4$} & \multicolumn{1}{c|}{$(39,9)$} & \multicolumn{1}{c|}{} & \multicolumn{1}{r|}{$\textbf{1.0}$} & \multicolumn{1}{c|}{$\textbf{(43,12,1)}$} \\ 
\multicolumn{1}{|c|}{} & \multicolumn{1}{r|}{$2.2$} & \multicolumn{1}{c|}{$(22,3)$} & \multicolumn{1}{c|}{} & \multicolumn{1}{r|}{$2.4$} & \multicolumn{1}{c|}{$(32,7)$} & \multicolumn{1}{c|}{} & \multicolumn{1}{r|}{$\textbf{1.4}$} & \multicolumn{1}{c|}{$\textbf{(38,10)}$} & \multicolumn{1}{c|}{} & \multicolumn{1}{r|}{$>5.0$} & \multicolumn{1}{c|}{$(41,14,1)$} \\ 
\cline{10-12}
\multicolumn{1}{|c|}{} & \multicolumn{1}{r|}{$>5.0$} & \multicolumn{1}{c|}{$(21,4)$} & \multicolumn{1}{c|}{} & \multicolumn{1}{r|}{$3.6$} & \multicolumn{1}{c|}{$(31,8)$} & \multicolumn{1}{c|}{} & \multicolumn{1}{r|}{$1.6$} & \multicolumn{1}{c|}{$(36,12)$} & \multicolumn{1}{c|}{$\textbf{58}$} & \multicolumn{1}{r|}{$\textbf{1.0}$} & \multicolumn{1}{c|}{$\textbf{(45,12,1)}$} \\ 
\cline{1-3}
\multicolumn{1}{|c|}{$26$} & \multicolumn{1}{r|}{$0.5$} & \multicolumn{1}{c|}{$(24,2)$} & \multicolumn{1}{c|}{} & \multicolumn{1}{r|}{$>5.0$} & \multicolumn{1}{c|}{$(30,9)$} & \multicolumn{1}{c|}{} & \multicolumn{1}{r|}{$>5.0$} & \multicolumn{1}{c|}{$(35,12,1)$} & \multicolumn{1}{c|}{} & \multicolumn{1}{r|}{$>5.0$} & \multicolumn{1}{c|}{$(43,14,1)$} \\ 
\cline{4-12}
\multicolumn{1}{|c|}{} & \multicolumn{1}{r|}{$0.7$} & \multicolumn{1}{c|}{$(23,3)$} & \multicolumn{1}{c|}{$\textbf{40}$} & \multicolumn{1}{r|}{$\textbf{0.4}$} & \multicolumn{1}{c|}{$\textbf{(34,6)}$} & \multicolumn{1}{c|}{$\textbf{49}$} & \multicolumn{1}{r|}{$0.2$} & \multicolumn{1}{c|}{$(40,9)$} & \multicolumn{1}{c|}{$59$} & \multicolumn{1}{r|}{$0.5$} & \multicolumn{1}{c|}{$(46,12,1)$} \\ 
\multicolumn{1}{|c|}{} & \multicolumn{1}{r|}{$>5.0$} & \multicolumn{1}{c|}{$(22,4)$} & \multicolumn{1}{c|}{} & \multicolumn{1}{r|}{$>5.0$} & \multicolumn{1}{c|}{$(32,8)$} & \multicolumn{1}{c|}{} & \multicolumn{1}{r|}{$\textbf{1.1}$} & \multicolumn{1}{c|}{$\textbf{(39,10)}$} & \multicolumn{1}{c|}{} & \multicolumn{1}{r|}{$0.9$} & \multicolumn{1}{c|}{$(45,13,1)$} \\ 
\cline{1-6}
\multicolumn{1}{|c|}{$27$} & \multicolumn{1}{r|}{$1.0$} & \multicolumn{1}{c|}{$(24,3)$} & \multicolumn{1}{c|}{$41$} & \multicolumn{1}{r|}{$0.1$} & \multicolumn{1}{c|}{$(35,6)$} & \multicolumn{1}{c|}{} & \multicolumn{1}{r|}{$1.4$} & \multicolumn{1}{c|}{$(37,12)$} & \multicolumn{1}{c|}{} & \multicolumn{1}{r|}{$>5.0$} & \multicolumn{1}{c|}{$(44,14,1)$} \\ 
\cline{10-12}
\multicolumn{1}{|c|}{} & \multicolumn{1}{r|}{$>5.0$} & \multicolumn{1}{c|}{$(23,4)$} & \multicolumn{1}{c|}{} & \multicolumn{1}{r|}{$0.6$} & \multicolumn{1}{c|}{$(34,7)$} & \multicolumn{1}{c|}{} & \multicolumn{1}{r|}{$>5.0$} & \multicolumn{1}{c|}{$(36,12,1)$} & \multicolumn{1}{c|}{$\textbf{60}$} & \multicolumn{1}{r|}{$0.1$} & \multicolumn{1}{c|}{$(48,12)$} \\ 
\cline{1-3}\cline{7-9}
\multicolumn{1}{|c|}{$28$} & \multicolumn{1}{r|}{$0.1$} & \multicolumn{1}{c|}{$(25,3)$} & \multicolumn{1}{c|}{} & \multicolumn{1}{r|}{$0.8$} & \multicolumn{1}{c|}{$(33,8)$} & \multicolumn{1}{c|}{$\textbf{50}$} & \multicolumn{1}{r|}{$0.1$} & \multicolumn{1}{c|}{$(41,9)$} & \multicolumn{1}{c|}{} & \multicolumn{1}{r|}{$\textbf{0.2}$} & \multicolumn{1}{c|}{$\textbf{(47,12,1)}$} \\ 
\multicolumn{1}{|c|}{} & \multicolumn{1}{r|}{$>5.0$} & \multicolumn{1}{c|}{$(24,4)$} & \multicolumn{1}{c|}{} & \multicolumn{1}{r|}{$>5.0$} & \multicolumn{1}{c|}{$(32,9)$} & \multicolumn{1}{c|}{} & \multicolumn{1}{r|}{$\textbf{0.7}$} & \multicolumn{1}{c|}{$\textbf{(40,10)}$} & \multicolumn{1}{c|}{} & \multicolumn{1}{r|}{$0.8$} & \multicolumn{1}{c|}{$(45,14,1)$} \\ 
\cline{1-3}
\multicolumn{1}{|c|}{$29$} & \multicolumn{1}{r|}{$1.6$} & \multicolumn{1}{c|}{$(25,4)$} & \multicolumn{1}{c|}{} & \multicolumn{1}{r|}{} & \multicolumn{1}{c|}{} & \multicolumn{1}{c|}{} & \multicolumn{1}{r|}{$1.6$} & \multicolumn{1}{c|}{$(38,12)$} & \multicolumn{1}{c|}{} & \multicolumn{1}{r|}{$2.9$} & \multicolumn{1}{c|}{$(44,15,1)$} \\ 
\multicolumn{1}{|c|}{} & \multicolumn{1}{r|}{$>5.0$} & \multicolumn{1}{c|}{$(24,5)$} & \multicolumn{1}{c|}{} & \multicolumn{1}{r|}{} & \multicolumn{1}{c|}{} & \multicolumn{1}{c|}{} & \multicolumn{1}{r|}{$>5.0$} & \multicolumn{1}{c|}{$(37,12,1)$} & \multicolumn{1}{c|}{} & \multicolumn{1}{r|}{$>5.0$} & \multicolumn{1}{c|}{$(43,16,1)$} \\ 
\hline
\end{tabular}
\caption{\label{table1}Table of structural transition points $\kappa_{cr}$, cf. Fig.~\ref{fig1}). Bold values mark anomalies of the 1st kind, where the inner shell changes by $2$ particles with increased screening ($N$ fixed). The screening values displayed are the critical values $(\pm0.05)$ up to which the configuration given in the 3rd column remains the ground state.}
\end{table}

\normalsize
\begin{table}
\centering
\begin{tabular}{|c|c|r|r|}
	\hline
	$N_1 \rightarrow N_2$ & configuration & $\kappa_{min}$ & $\kappa_{max}$\\
	\hline
	$11 \rightarrow 12$ & $(10,1)\rightarrow (12)$ 		  &$ 1.9038 $&$ 4.0567$\\ 
	\hline
	$20\rightarrow 21 $&$ (18,2)\rightarrow (20,1)$ 	  &$ 1.0762 $&$ 1.1906$\\ 
	\hline
	$25\rightarrow 26 $&$ (22,3)\rightarrow (24,2)$  	  &$ 0.2544 $&$ 0.5049$\\ 
	\hline
	$26\rightarrow 27 $&$ (22,4)\rightarrow (24,3)$ 	  &$ 0.7287 $&$ 1.0412$\\ 
	\hline
	$30\rightarrow 31 $&$ (24,6)\rightarrow (27,4)$ 	  &$ 1.4866 $&$ 1.5623$\\ 
	\hline
	$30\rightarrow 31 $&$ (24,6)\rightarrow (26,5)$ 	  &$ 1.5623 $&$ 1.6142$\\ 
	\hline
	$33\rightarrow 34 $&$ (28,5)\rightarrow (30,4)$ 	  &$ 0.2012 $&$ 0.2450$\\ 
	\hline
	$34\rightarrow 35 $&$ (28,6)\rightarrow (30,5)$ 	  &$ 0.2450 $&$ 0.3034$\\ 
	\hline
	$35\rightarrow 36 $&$ (28,7)\rightarrow (30,6)$ 	  &$ 3.1665 $&$ 3.6133$\\ 
	\hline
	$36\rightarrow 37 $&$ (28,8)\rightarrow (30,7)$  	  &$ 3.6133 $&$ 4.1646$\\ 
	\hline
	$37\rightarrow 38 $&$ (30,7)\rightarrow (32,6)$ 	  &$ 1.6679 $&$ 2.0283$\\ 
	\hline
	$38\rightarrow 39 $&$ (30,8)\rightarrow (32,7)$ 	  &$ 2.0283 $&$ 2.4396$\\ 
	\hline
	$39\rightarrow 40 $&$ (32,7)\rightarrow (34,6)$ 	  &$ 0.2223 $&$ 0.4179$\\ 
	\hline
	$39\rightarrow 40 $&$ (31,9)\rightarrow (32,8)$ 	  &$ 3.6120 $&$ >5.0000$\\ 
	\hline
	$40\rightarrow 41 $&$ (32,8)\rightarrow (34,7)$ 	  &$ 0.4179 $&$ 0.5521$\\ 
	\hline
	$41\rightarrow 42 $&$ (32,9)\rightarrow (34,8)$ 	  &$ 0.8329 $&$ 1.1372$\\ 
	\hline
	$42\rightarrow 43 $&$ (32,10)\rightarrow (34,9)$	  &$ 1.8473 $&$ 3.6391$\\ 
	\hline
	$49\rightarrow 50 $&$ (36,12,1)\rightarrow (38,12)$ 	  &$ 1.3753 $&$ 1.5634$\\ 
	\hline
	$55\rightarrow 56 $&$ (42,12,1)\rightarrow (44,12)$	  &$ 0.4964 $&$ 0.5150$\\ 
	\hline
	$59\rightarrow 60 $&$ (47,12,1)\rightarrow (48,12)$ 	  &$ 0.0000 $&$ 0.1024$\\ 
	\hline
\end{tabular}
\caption{\label{table2}Table of anomalies of the 2nd kind. Left column shows the change of the total particle number by one and column two the associated configuration change. The third and fourth column give the range of screening parameters where this transition occurs.}
\end{table}

\subsection{\label{ssec_ano3} Anomalies of the third kind: Reentrant shell transition upon increase of $\kappa$}
Finally, there is a the third kind of anomalous behavior which deviates from the ``normal'' shell filling trend of increased populations of the inner shells upon increase of $\kappa$ at a constant $N$. This tendency is never violated in the considered range of particle numbers, $11 \le N \le 60$, and for $\kappa  \le 5$. Since $\kappa=5$ corresponds to a pair interaction of very short-range one might expect that further increase of $\kappa$ will not change the cluster structure qualitatively. To verify whether this is the case we performed, for all $N$, additional calculations for an even larger screening,  $\kappa = 20.0$, cf. Fig.~\ref{fig1}. In most cases there is, indeed, no further change of the ground state configuration compared to $\kappa=5$, as expected. For three particle numbers, $N=44,57,58$, the ground state configuration still changes in the ``normal'' way such that one particle is relocated from the outer shell to the inner shell. 

\begin{table}
\centering
\begin{tabular}{|c|c|r|c|r|c|}
\hline
$N$ & conf. $1$ & $\kappa_{cr1}$ & conf. $2$ & $\kappa_{cr2}$ & conf. $3$ \\
\hline
$35$ & $(29,6)$ & $3.2$ & $(28,7)$ & $6.84$ & $(29,6)$ \\
\hline
$36$ & $(30,6)$ & $3.6$ & $(29,7)$ & $8.76$ & $(30,6)$ \\
\hline
$37$ & $(30,7)$ & $4.2$ & $(29,8)$ & $6.91$ & $(30,7)$ \\
\hline
$39$ & $(31,8)$ & $3.6$ & $(30,9)$ & $13.40$ & $(31,8)$ \\
\hline
$54$ & $(41,12,1)$ & $1.68$ & $(40,13,1)$ & $5.04$ &  \\
 &  & $5.04$ & $(39,14,1)$ & $15.02$ & $(41,12,1)$ \\
\hline
\end{tabular}
\caption{\label{table4}Table of anomalies of the 3rd kind. The 1st configuration is the ground state configuration up to the critical screening $\kappa_{cr_1}$, then the ground state configuration changes in the standard way by adding a particle on a inner shell. This configuration then is the ground state up to the critical screening $\kappa_{cr_2}$, at which the cluster changes its ground state configuration back to the one it had a lower screening ({\em anomalie of the 3rd kind}).}
\end{table}

However, there are six remarkable cases which violate this trend: $N=35,36,37,39,54$ and $N=43$. Consider first the total particle number $N=43$. This case is interesting because it is the only case where the central configuration $(11,1)$ is part of the ground state, apart from the cluster with $12$ at screenings $\kappa \ge 4.1$. This arrangement does otherwise not occur because the clusters prefer the platonic body with $12$ particles on the inner shell (closed shell configuration). Here the configuration $(31,11,1)$ becomes the ground state at $\kappa=17.4$ and remains the ground state for larger screening.

In the other five cases we observed, at $\kappa = 20.0$, several stationary states which differed only very little in their energies. We, therefore, performed molecular dynamics simulations using separately each of these states  as an input at $\kappa = 20.0$ and then decreased the screening slightly, letting the system relax into a new stationary state, often with the same configuration and symmetry. This way we could be certain to follow all metastable states and independently record their energy dependence on $\kappa$. The above five cases fall into two groups which differ with respect to the cluster symmetry. For the first, i.e. $N=35,36,37,39$, the cluster decreases the number of particles on the inner shell when the screening is increased between $\kappa = 5.0$ to $20.0$. The resulting new ground state configuration contains again a platonic body on the inner shell. Allowing for such a highly symmetric configuration here turns out to be energetically more favorable compared to the previous shell configuration or a simple increase of the number of particles on the inner shell. 

\begin{figure}
	\centering
\begin{minipage}[t]{7.5cm}
	\includegraphics[width=7.45cm]{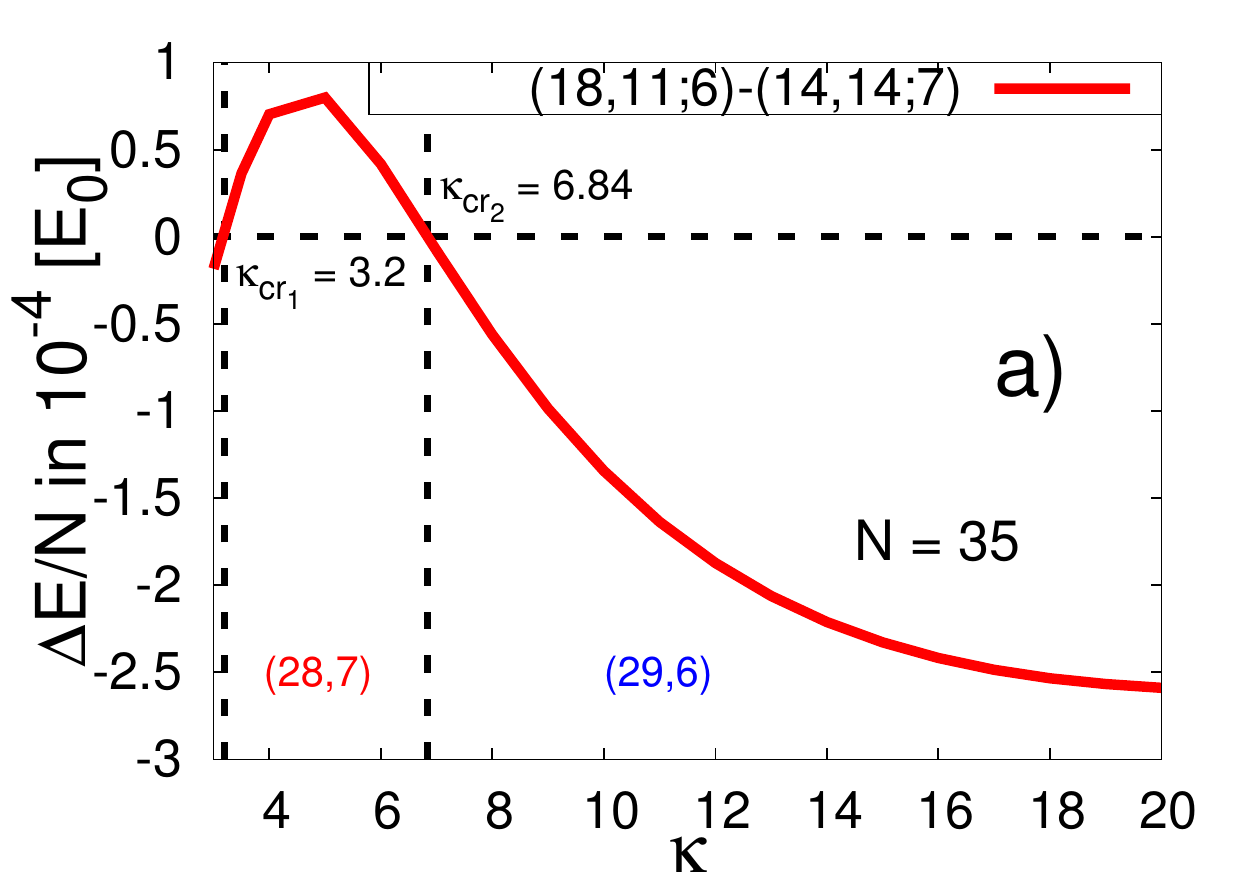}
	\includegraphics[width=7.45cm]{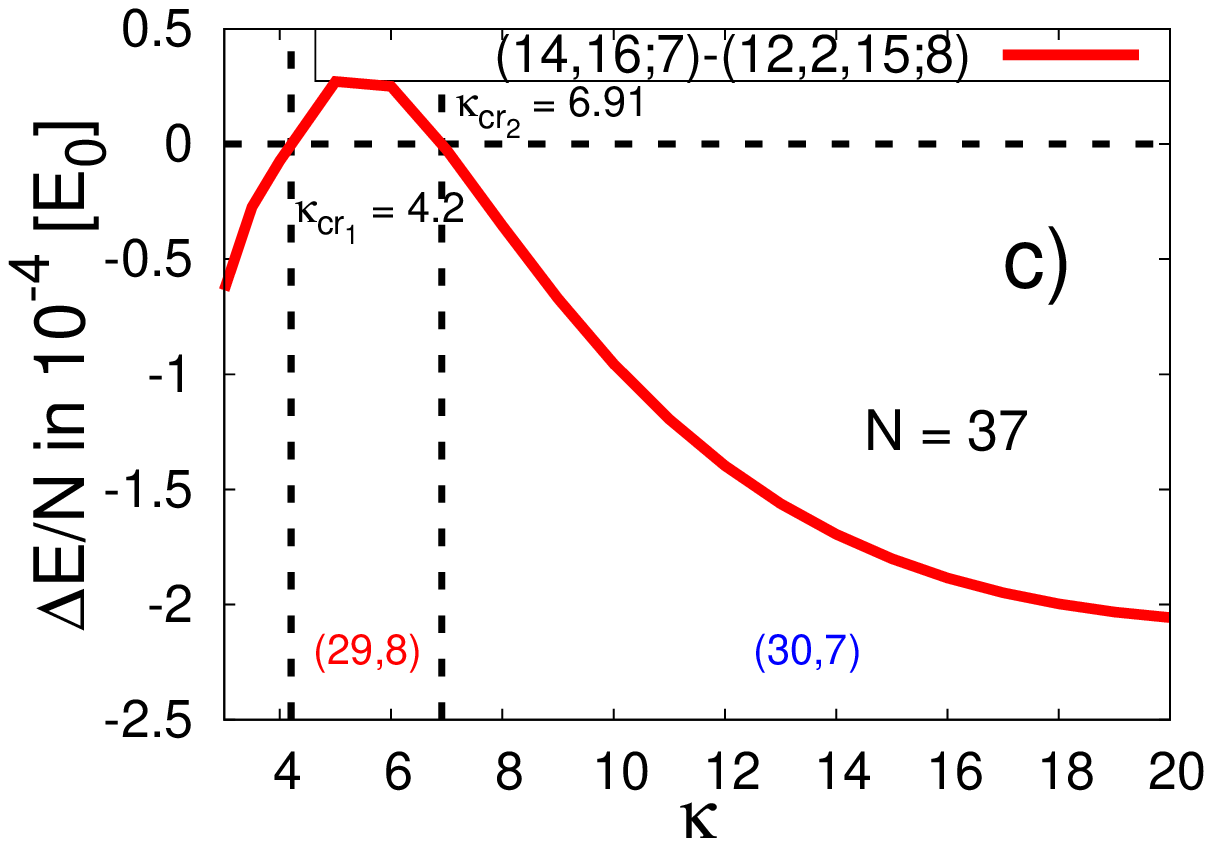}
\end{minipage}
\begin{minipage}[t]{7.5cm}
	\includegraphics[width=7.45cm]{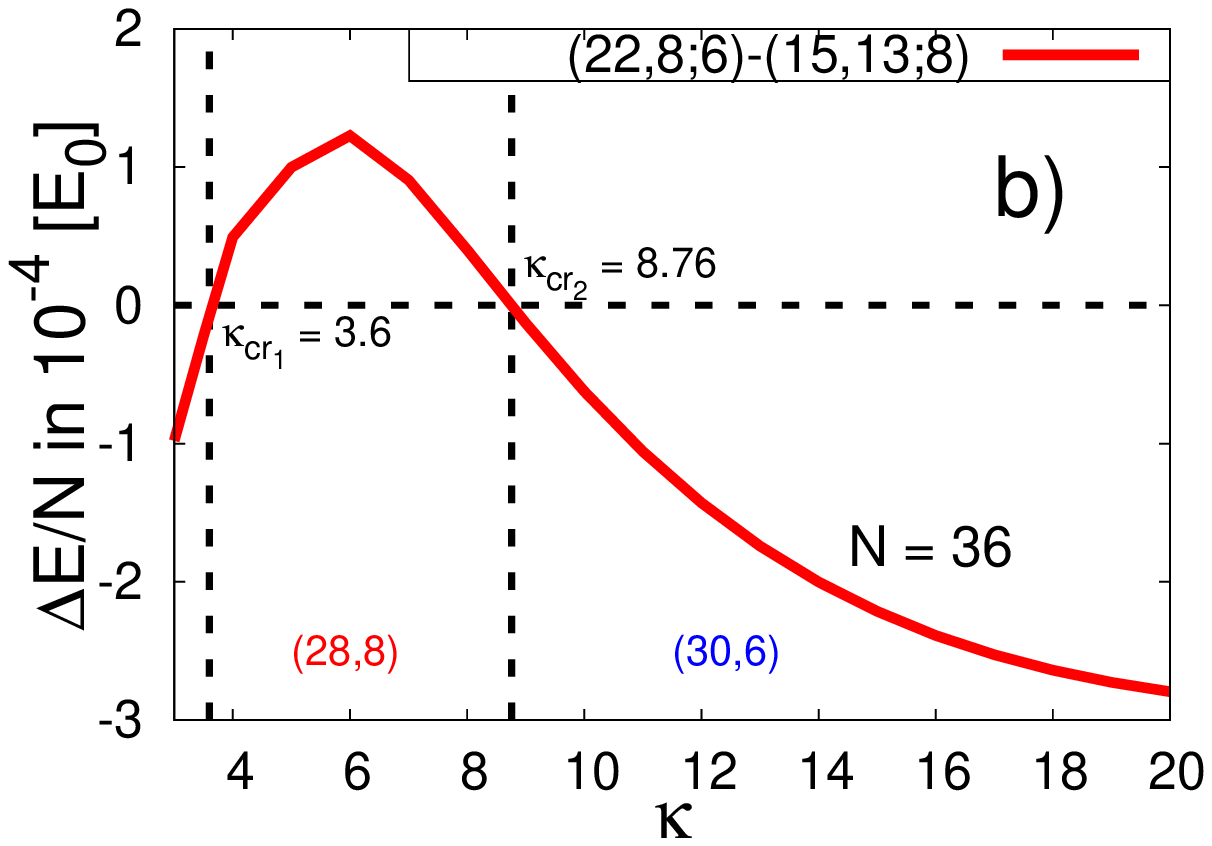}
	\includegraphics[width=7.45cm]{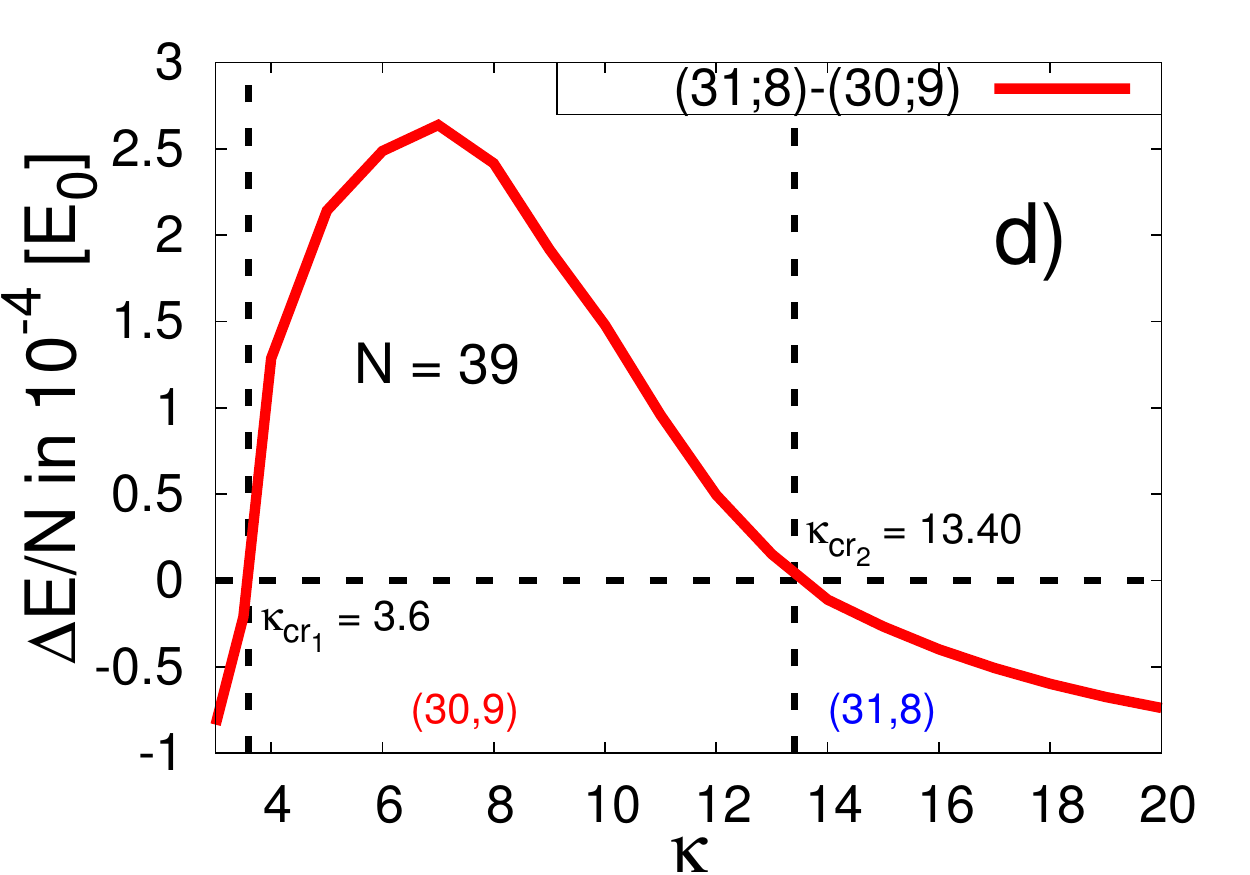}
\end{minipage}
	\caption{\label{fig7} (colour online) Reentrant shell configuration changes for $N=35$ (top left), $N=36$ (top right), $N=37$ (bottom left) and $N=39$ (bottom right). When $\kappa$ is increased, at $\kappa_{cr1}$ one particle moves towards the center and, at $\kappa_{cr2}$, one particle returns to the outer shell restoring the former ground state configuration. The critical values of $\kappa$ are indicated by the vertical dashed lines. The solid red line shows the energy difference of the two configurations which has two zeroes. The legend shows the shell configurations, including the splitting of the outer shell into subshells, given by the numbers in square brackets.}
\end{figure}

\begin{figure}[h]
	\centering
	\includegraphics[width=8.5cm]{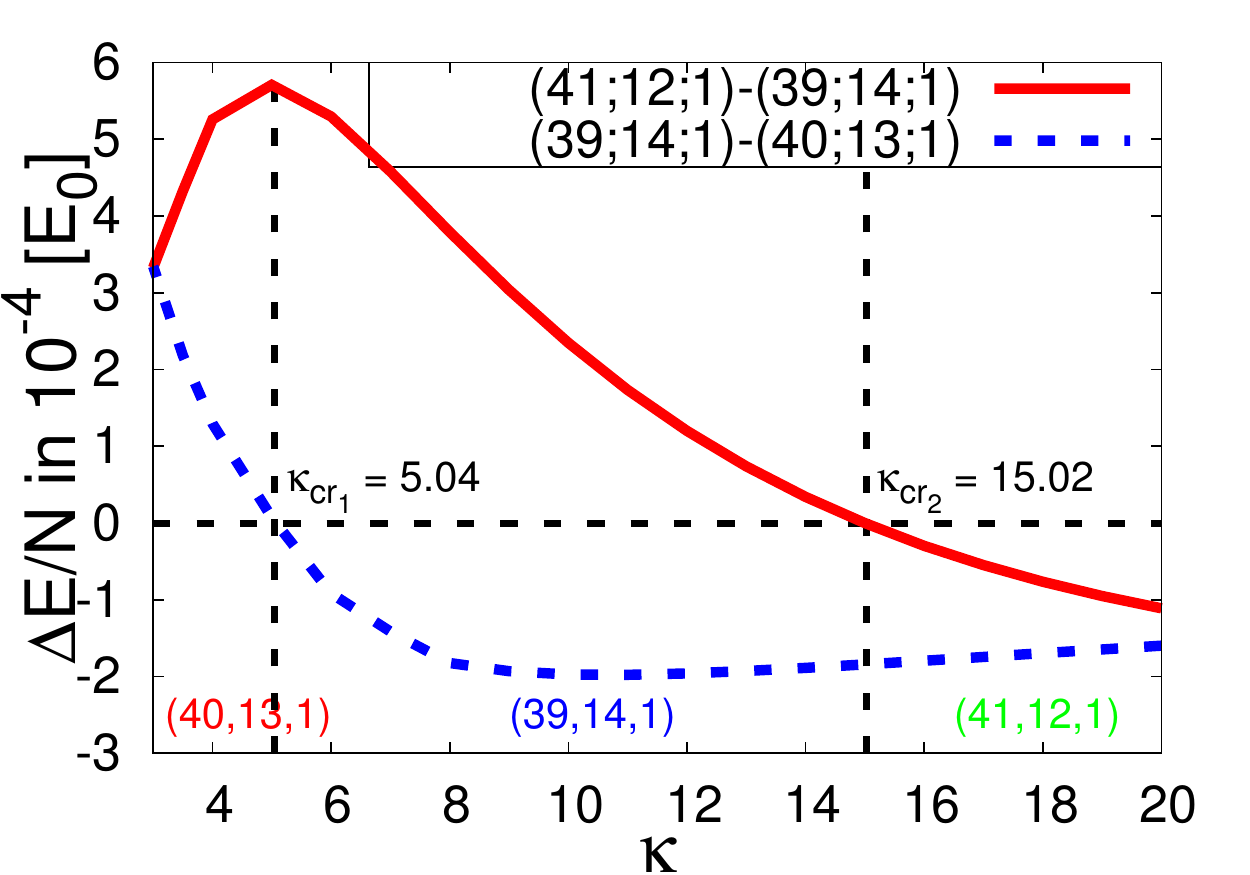}
	\caption{\label{fig6} (colour online) The energy differences of all states which become the ground state in the screening range $4.0 \le \kappa \le 20.0$ for the particle number $N=54$. In this case increase of $\kappa$ leads to a ground state with fewer particles on the second shell, returning to a ground statea configuration, that already existed at lower screenings. In this case an ``Anomaly of the third kind'' is observed.}
\end{figure}

Consider, for example, the cluster $N=39$, cf. Fig.~\ref{fig7}.d. Here, at $\kappa=5$ the ground state is $(30,9)$ until, at $\kappa_{cr}=13.40$, the configuration $(31,8)$ with one particle less on the inner shell becomes the ground state. 
\\
The particle number reduction on inner shells is sometimes accompanied by another trend: with increasing $\kappa$ shells tend to split into subshells with close radii, as was already observed in Refs.~\cite{baum07,apolinario_phd}. This is observed e.g. for  $N=35$, cf. Fig.~\ref{fig7}.a. Here the configuration $(28,7)$ which is the ground state at $\kappa=5$ has in fact two subshells each containing $14$ particles which we will denote $([14,14],7)$. The radii of the two subshells differ only slightly, $R_{2,1} = 0.947$ and $R_{2,2} = 0.875$, respectively, while the inner shell radius is $R_1 = 0.426$, clearly distinguishable from the outer shell. At $\kappa_{cr}=6.84$ we observe a transition $([14,14],7) \longrightarrow ([18,11],6)$, i.e. one particle from the inner shell moves outward and, in addition, three particles from the inner subshell move to the outer subshell. 

Similar behavior is observed for  $N=36$, cf. Fig.~\ref{fig7}.b. Here the configuration $([15,13],8)$ is the ground state at $\kappa=5$. At $\kappa_{cr}=6.84$ we observe a transition $([15,13],8) \longrightarrow ([22,8],6)$, where the inner shell loses two particles and, in addition, the inner subshell transfers 5 particles to the outer subshell.
Analogously, for  $N=37$, cf. Fig.~\ref{fig7}.c, the configuration $([12,2,15],8)$ is the ground state at $\kappa=5$. At $\kappa_{cr}=6.91$ we observe a transition $([12,2,15],8) \longrightarrow ([14,16],7)$, where the inner shell loses one particle.

The cluster with $N=54$ shows a similar behavior, cf. Fig.~\ref{fig6}. Here, first the second shell popluation increases by one, at $\kappa_{cr1}=5.04$, according to $(40,13,1)\longrightarrow (39,14,1)$. Further increase of screening makes a third configuration more favorable which has even two particles less on the second shell: the configuration $(41,12,1)$ becomes the ground state again at $\kappa_{cr2}=15.02$ which again is a consequence of the high symmetry (closed shell configuration).

Finally, particularly interesting behavior is observed for all mentioned $N=35, 36, 37, 39, 54$, if a larger range of screening is considered, cf. Fig.~\ref{fig7} and Fig.~\ref{fig6}. Here, for $\kappa \ge 1.7$, there are two states which become the ground state. At a first critical value $\kappa_{cr1}$ one particle moves to the inner shell until at  $\kappa_{cr2}$ this transition is reversed (the only difference for $N=54$ is that there is an additional ground state configuration between these two critical screenings): one particle moves outward and the original configuration with fewer particles on the inner shell is restored which remains the ground state for all larger values of $\kappa$. This contradiction to the general trend (of increasing the inner shell population with increased screening), together with the reappearance of a ground state configuration, will be called ``anomaly of the third kind''. The complete set of these cases can be found in Tab.~\ref{table4} with the exact critical screening parameters for the ground state configuration changes. The reentrance of these ground state configurations at large screening are in all cases not different in their symmetry compared to the ground state configurations below $\kappa_{cr1}$, they have the same number of nearest neighbors and same shape of the Voronoi cells, with only their length scale strongly reduced due to the weaker interaction force. 

In general, the restored ground state configurations consist of platonic bodies on the inner shell, except for the case of $N=37$. Here, the ground state configuration changes from $(30,7)$ to $(29,8)$ at a screening value of $\kappa = 4.2$ and back to $(30,7)$ at a screening value of $\kappa = 6.91$. The $7$ particles on the inner shell are not a platonic body and, as one can see from Fig. \ref{fig1}, it is not a common configuration compared to $6$ or $8$ particles in the center. Nevertheless, this can be understood by looking at the outer shell. The $30$ particles are placed on the edges of an icosaeder which results in a highly symmetric configuration for the outer shell.

\section{\label{sec4}Conclusions}
The goal of this paper was to present, for the first time, a detailed analysis of the ground state shell configurations of Yukawa clusters in a parabolic shperical confinement over a broad range of screening parameters $\kappa\le 20$. This allowed us to analyze the structural transitions occuring when the pair interaction changes from long range, in the Coulomb case, to short range, at the largest values of $\kappa$. Focusing on a finite range of particle numbers, $11\le N\le 60$ we presented a complete overview on all existing changes of the shell configurations for $\kappa\le 5$. For larger $\kappa$ we also noted the cases where the configurations are different at $\kappa=5$ and $\kappa=20$ (we cannot rule out that, in this range, there occur transtions in addition to those given). The general trend observed earlier \cite{bonitz06} was confirmed: with increased screening, more particles tend to populate the inner shell(s) of the cluster giving rise to an average density profile which decreases increasingly fast towards the edge \cite{henning06,henning07}. There are, however, three non-trivial deviations (``anomalies'') from this shell filling sequence which were analyzed: 1.) upon $\kappa$ increase two particles move to (one of) the inner shell(s) at once. 2.) when the particle number is increased by one, at a fixed $\kappa$, one particle moves from the inner to the outer shell and 3.) at very large $\kappa$ there exist cases of reentrent shell fillings: one particle returns from the inner to the outer shell. These anomalies are, in most cases, dictated by symmetry properties of the corresponding state which allow to lower the total energy.

Our results are expected to be useful also for experiments with dusty plasmas and allow us to predict interesting parameter ranges which give information on the effect of symmetry on the structure of mesoscopic systems. In current experiments on spherical dust crystals performed at the Universities Kiel and Greifswald \cite{arp04} typical values of the screening parameter are in the range of $0.6 \le \kappa \le 1.6$. While this gives access only to a small part of the analyzed parameters where no reentrant shell fillings (third anomaly) occur, still the first two effects should be observable.
While the experiments on small clusters do not necessarily reveal the ground state configurations,  since often metastable states occur with a higher probability \cite{block08,kaehlert08} the prediction of parameters where two states have the same energy is of practical interest for the analysis of potential energy barriers and intershell transitions. 

\section{\label{sec5}Acknowledgements}
The authors would like to thank A. Melzer and D. Block for stimulating discussions. This work is supported by the Deutsche Forschungsgemeinschaft via SFB-TR 24, grants A5 and A7 and by the U.S. Department of Energy award DE-FG02-07ER54946.

\end{document}